\documentclass[usenatbib,useAMS]{mn2e}
\usepackage{graphicx}
\usepackage{epstopdf}
\usepackage{ctable}
\usepackage{url}
\usepackage{times}

\newcommand{\acknowledgments}{\begin{small}\section*{Acknowledgments}\end{small}}

\title[Blind $^{12}$CO search in HS1700]{A blind CO detection of a Distant Red Galaxy in the HS1700+64 proto-cluster.}

\author[S.~C.\ Chapman et al.]{
\parbox[t]{\textwidth}{
S.\,C.\ Chapman$^{1}$\thanks{scott.chapman@dal.ca},
F.\ Bertoldi,$^{2}$
Ian Smail,$^{3}$
A.\,W.\ Blain,$^{4}$
J.\,E.\ Geach,$^{5}$
M.\ Gurwell,$^{6}$\ \ \ \ \ 
R.\,J.\ Ivison,$^{7}$
G.\,R.\ Petitpas,$^{6}$ 
N.\ Reddy,$^{8}$
C.\,C.\ Steidel$^{9}$}
\vspace*{6pt} \\
$^1$Department of Physics and Atmospheric Science, Dalhousie University, Halifax, NS, B3H 4R2, Canada\\
$^2$ Argelander-Institute of Astronomy, Bonn University, Auf dem Hugel 71, D-53121 Bonn, Germany\\
$^{3}$Institute for Computational Cosmology, Department of Physics, Durham University, South Road, Durham DH1 3LE\\
$^{4}$Physics and Astronomy, University of Leicester, University Road Leicester, LE1 7RH\\
$^{5}$Centre for Astrophysics, Science \& Technology Research
Institute, University of Hertfordshire, Hatfield AL10 9AB, UK\\
$^{6}$Harvard-Smithsonian Center for Astrophysics, 60 Garden Street, Cambridge, MA 02138\\
$^7$Institute for Astronomy, University of Edinburgh, Royal Observatory, Blackford Hill, Edinburgh, EH9 3HJ, UK\\
$^8$Department of Physics and Astronomy, UC Riverside, 900 University
Avenue, Riverside, CA 92521\\
$^{9}$Cahill Center for Astronomy and Astrophysics, California Institute of Technology, MS 249-17, Pasadena, CA 91125, USA
}
\usepackage{hyperref}

\begin{document}

\date{Submitted for publication in MNRAS}

\pagerange{\pageref{firstpage}--\pageref{lastpage}} \pubyear{2013}

\maketitle

\label{firstpage}

\begin{abstract}
We report the blind  detection of  $^{12}$CO emission from a Distant Red Galaxy, HS1700.DRG55.    
We have used the
IRAM PdBI-WIDEX, with its 3.6\,GHz of instantaneous
dual-polarization bandwidth, to target $^{12}$CO(3--2) from galaxies lying in the proto-cluster at $z=2.300$ in the field HS1700+64.  
If indeed this line in DRG55  is $^{12}$CO(3--2), its detection at 104.9GHz  indicates a $z_{\rm CO}$=2.296. 
None of the other eight known $z\sim2.30$ proto-cluster galaxies lying within the primary beam (PB) are detected in $^{12}$CO, although the limits are $\sim2\times$ worse towards the edge of the PB where several lie.
The optical/near-IR magnitudes of DRG55 ($R_{AB}>27$, $K_{\rm AB}=22.3$) mean that optical spectroscopic redshifts are difficult with 10m-class telescopes, but near-IR redshifts would be feasible. The 24$\mu$m-implied SFR (210 M$_\odot$ yr$^{-1}$), stellar mass ($\sim10^{11}$ M$_\odot$)  and $^{12}$CO line luminosity ($3.6\times10^{10}$ K km s$^{-1}$ pc$^2$)  are comparable to other {\it normal} $^{12}$CO-detected star forming galaxies in the literature, although the galaxy is some $\sim$2 mag ($\sim6\times$) fainter in the rest-frame UV  than $^{12}$CO-detected galaxies at $z>2$. 
The detection of DRG55 in $^{12}$CO complements three other $^{12}$CO detected UV-bright galaxies in this proto-cluster from previous studies, 
and suggests that many optically faint galaxies in the proto-cluster may host substantial molecular gas reservoirs, and a full blind 
census of $^{12}$CO in this overdense environment is warranted. 
\end{abstract}

\begin{keywords}
galaxies: abundances -- galaxies: high-redshift --  submillimeter: galaxies.
\end{keywords}

\section{Introduction} \label{S:intro}

Massive galaxy clusters at $z$$>$1 show a reversal in the star-formation density relation such that there is an enhancement of activity in the highest density regions  (e.g.\ Elbaz et al.\ 2007,2011), with observations revealing increasing levels of activity even in the cluster cores 
(Cooper et al.\ 2008; Chapman et al.\ 2009; Hilton et al.\ 2010; Tran et al.\ 2010; Smail et al.\ 2014). 
By contrast, the gas properties of galaxies in distant clusters remain poorly constrained, despite the fact that they may well elucidate the mechanisms for the increasing SFRs in cores of massive clusters.
Studies of the molecular medium in distant galaxies provide key diagnostics about the evolutionary state of galaxies in the high-redshift universe. To date, over two hundred high-redshift galaxies have been detected in $^{12}$CO emission, the main tracer for molecular gas (e.g., Bothwell et al.\ 2013, Tacconi et al.\ 2013). Essentially all these detections were obtained by targeted observations of galaxies that have been pre-selected through their star forming properties, e.g.\ Ultra-Violet (UV) or far-infrared (FIR) emission.
However, the gas supplies, star formation efficiencies, and starburst modes  (merger driven versus quiescent disk) may vary strongly as a function of their local density, and comparing the $^{12}$CO properties of galaxies in proto-clusters to the field should elucidate the stronger evolution seen in overdense regions.

In this Letter, we describe a blind 3-mm survey for redshifted $^{12}$CO(3--2) molecular gas from $z\sim2.30$ galaxies  in  HS1700+64,
a proto-cluster with a
 redshift-space galaxy overdensity of $\delta^z_g\sim7$ and an estimated matter overdensity 
 which indicates that it will virialize by $z$$\sim$0 with a mass scale of a rich galaxy cluster, $\sim14\times10^{15}$ M$_\odot$ (Steidel et al.\ 2005). 
Low-J $^{12}$CO transitions are collisionally excited by H$_2$ at low temperatures providing a good census of the star forming gas  in a range of galaxy types/luminosities 
(e.g., Ivison et al.\ 2011; Riechers et al.\ 2011; Harris et al.\ 2012). 
In the Tacconi et al.\ (2013) study of $^{12}$CO(3--2) in {\it normal} star forming galaxies, six of the 19  targets lying at $z>2$ are in this HS1700+64 field, with four identified to UV-bright proto-cluster members at $z\sim2.3$, providing a reference sample for our study.
The 3-mm band provides a reasonably large field of view on the IRAM-Plateau de Bure Interferometer  ($\sim50''$ primary beam,  $\sim$0.4\ Mpc diameter at $z=2.3$), with 
the  strong clustering of galaxies in angular scales and  along the line of sight effectively increasing the number of galaxies observed within a single pointing with the PdBI (see also Tadaki et al.\ 2014). 
We use cosmological parameters $\Omega_m$ = 0.3, $\Lambda_0$ = 0.7, and H$_0$ = 70 km s$^{-1}$ Mpc$^{-1}$ throughout the paper; at $z=2.30$, this corresponds to an angular scale of 7.9 kpc arcsec$^{-1}$. 

\begin{figure*}
\centering
\includegraphics[width=5.8cm,angle=0]{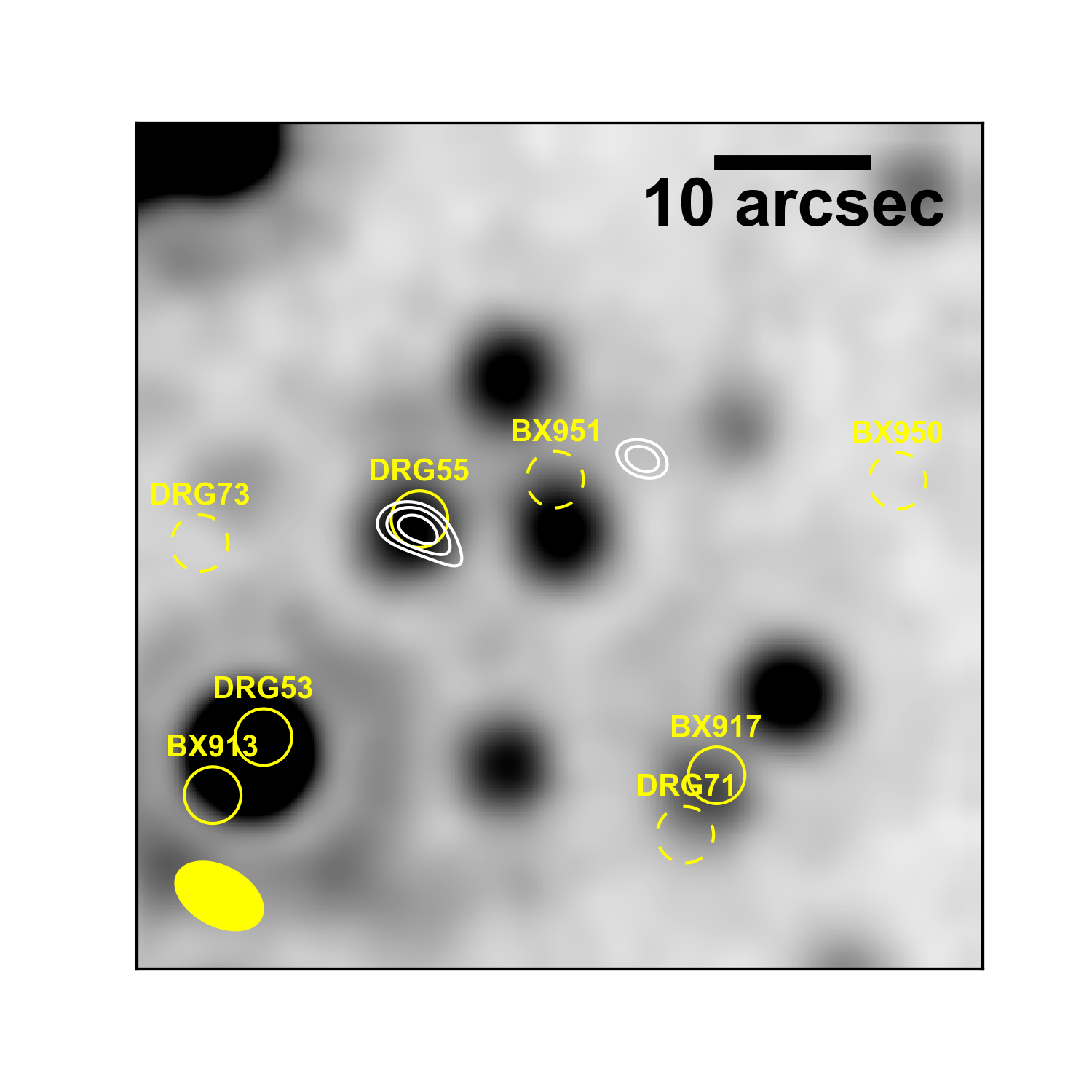}
\includegraphics[width=5.9cm,angle=0]{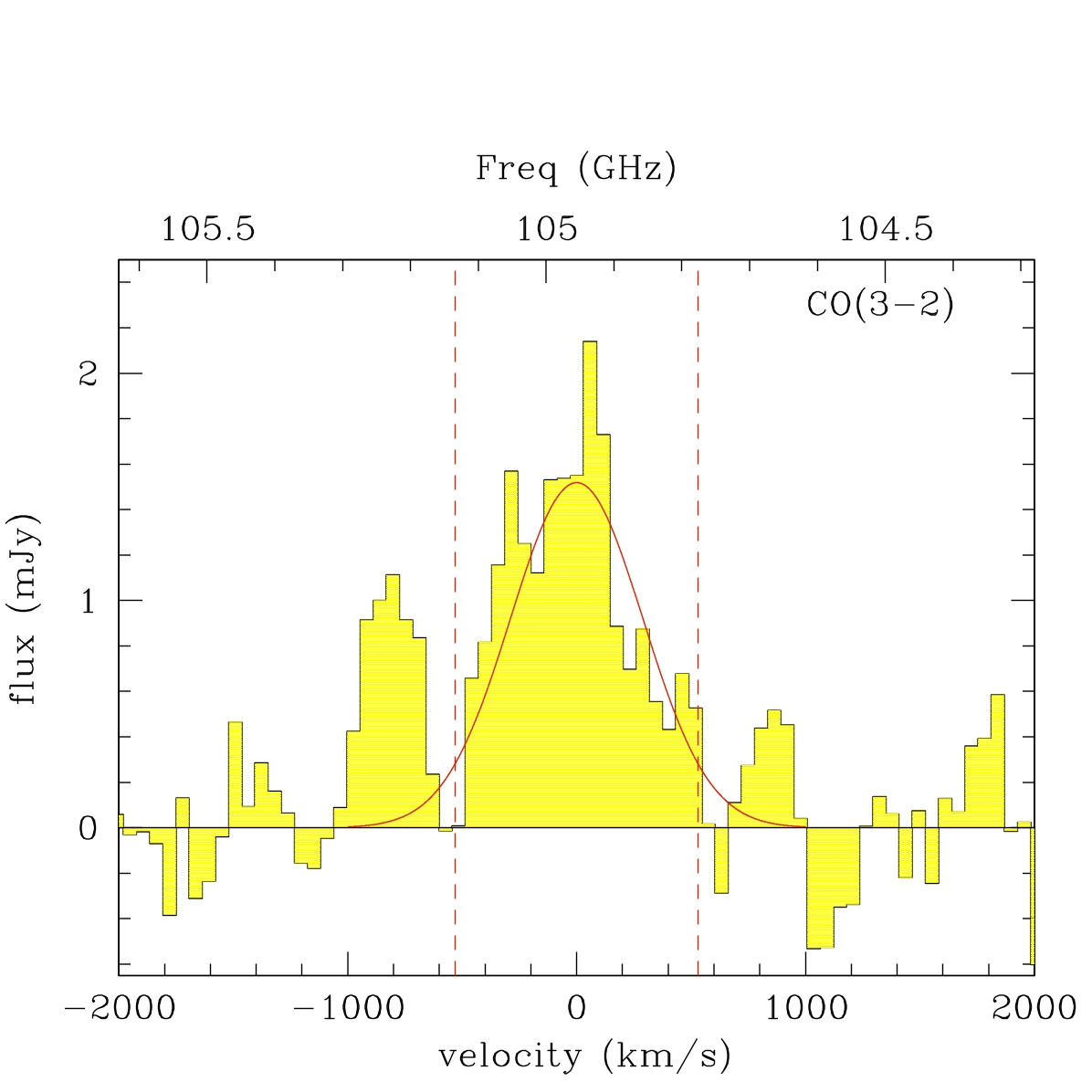}
\includegraphics[width=5.8cm,angle=0]{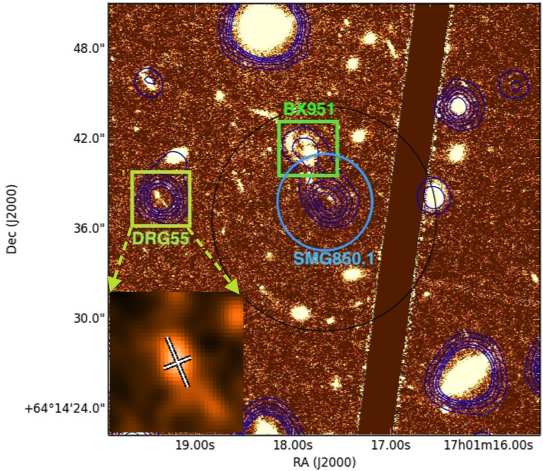}
\caption{{\bf Left:}  A {\it Spitzer}-MIPS 24$\mu$m image with the $^{12}$CO channel map at $z$$\sim$2.30 from the HS1700+64 PdBI pointing overlaid, constructed by summing channels from 104.76 to 105.15 GHz 
($\sim$1100 km/s), showing a 4.8$\sigma$ detection of 1.1$\pm$0.2\,mJy at the position of DRG55 (contours), coincident with a 160$\mu$Jy
24$\mu$m source. The primary beam of PdBI (48$''$) is larger than the field shown. 
{\bf Middle:} The extracted 1D spectrum shows a well detected CO(3-2) line 
with a continuum RMS=0.23\,mJy. A gaussian fit to the line suggests $z_{\rm CO}=2.296$. The velocity limits of the channel map (left) are shown as dashed lines.
{\bf Right:} An {\it HST} ACS F814W image (details in Peter et al.\ 2007) 
 centered on the PdBI observation. The DRG55 is identified, along with a UV-luminous BX galaxy and an S$_{\rm 850 \mu m}$=19.1\,mJy SMG. Blue contours show IRAC 4.5$\mu$m emission. DRG55 is identified with a faint galaxy with {\it tadpole} morphology, also shown in the 1.5$''$ field inset with a Gaussian fit to the galaxy of FWHM 0.7$''$ $\times$ 0.4$''$.
}
\label{spectra} 
\end{figure*}

\begin{figure*}
\centering
\includegraphics[width=5.9cm,angle=0]{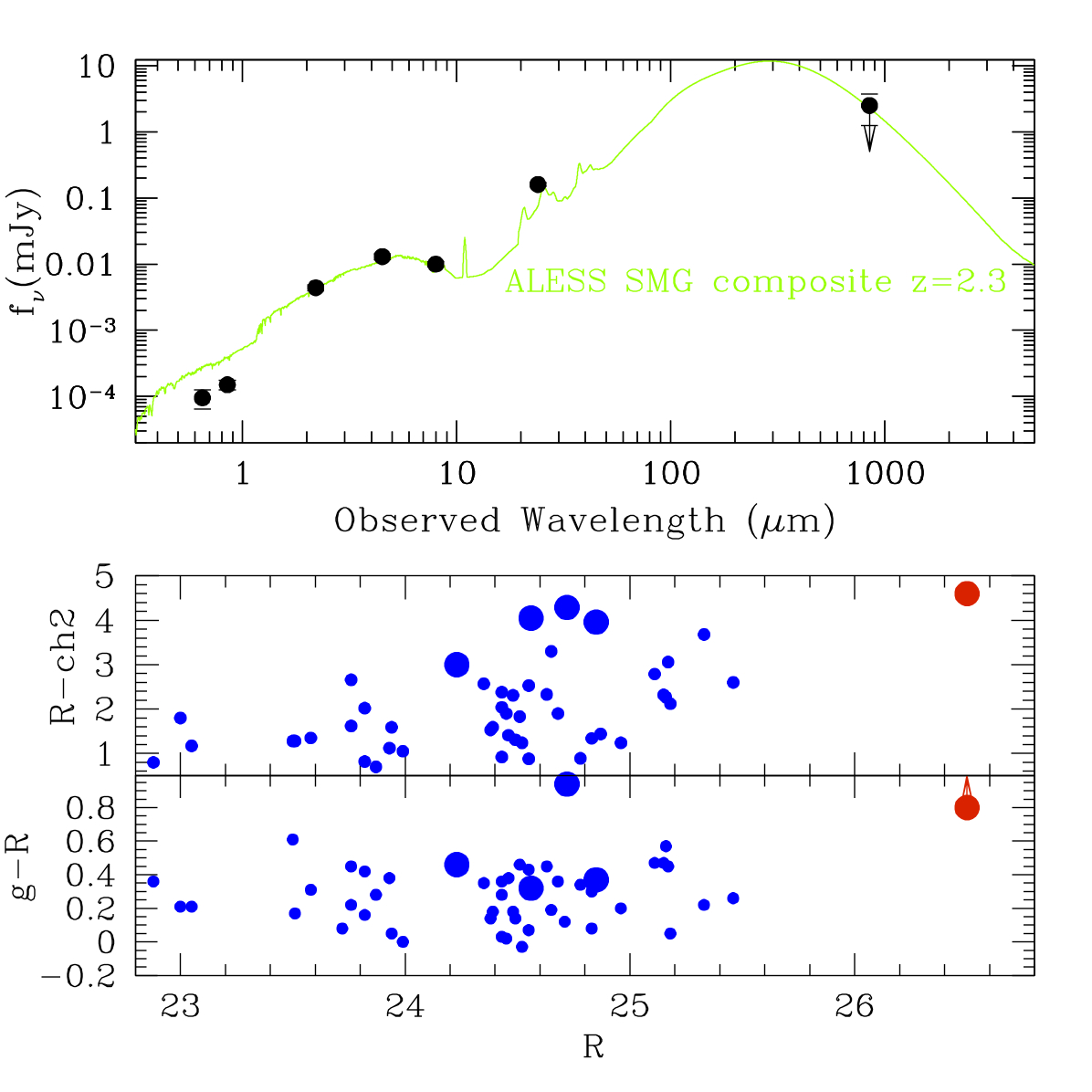}
\includegraphics[width=5.8cm,angle=0]{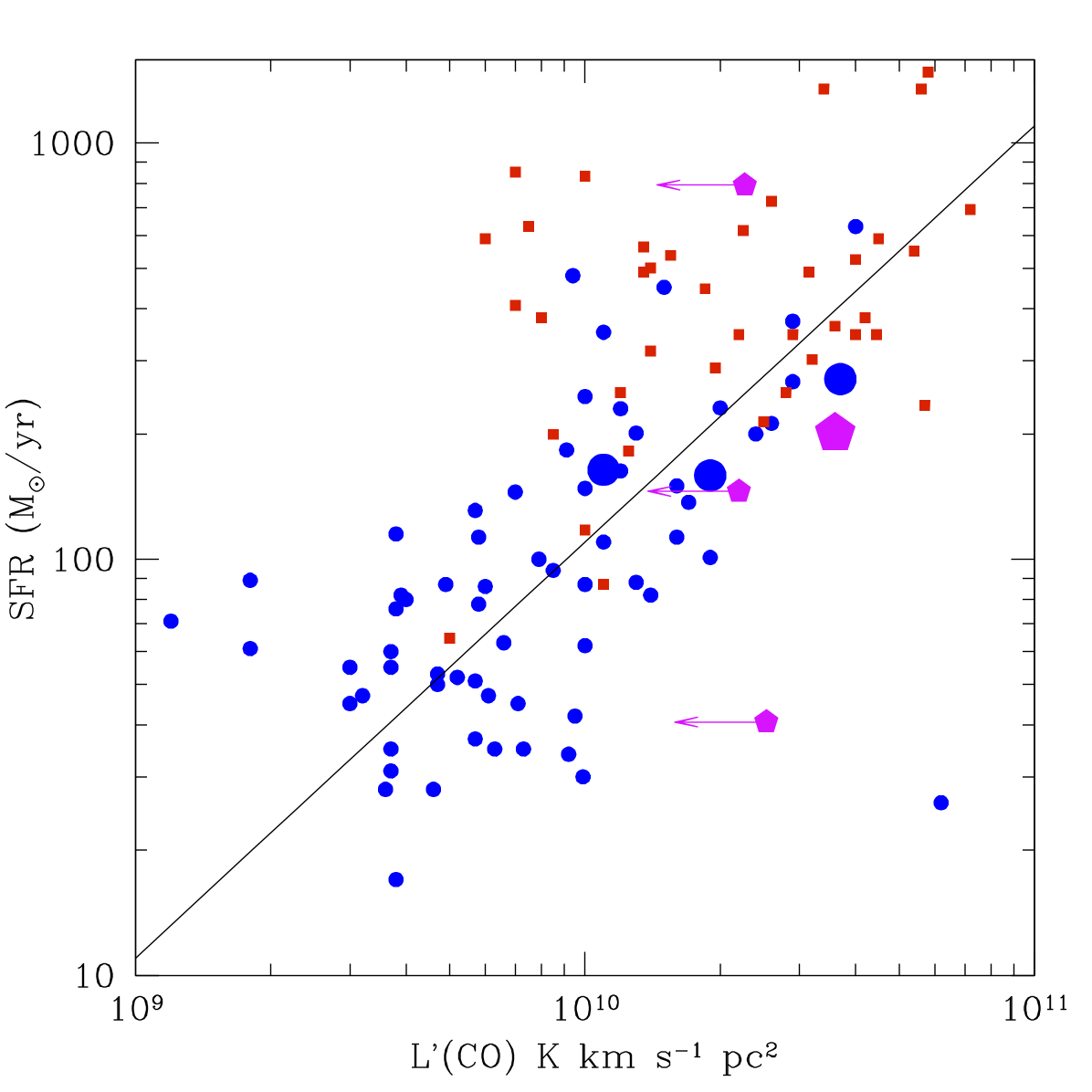}
\includegraphics[width=5.85cm,angle=0]{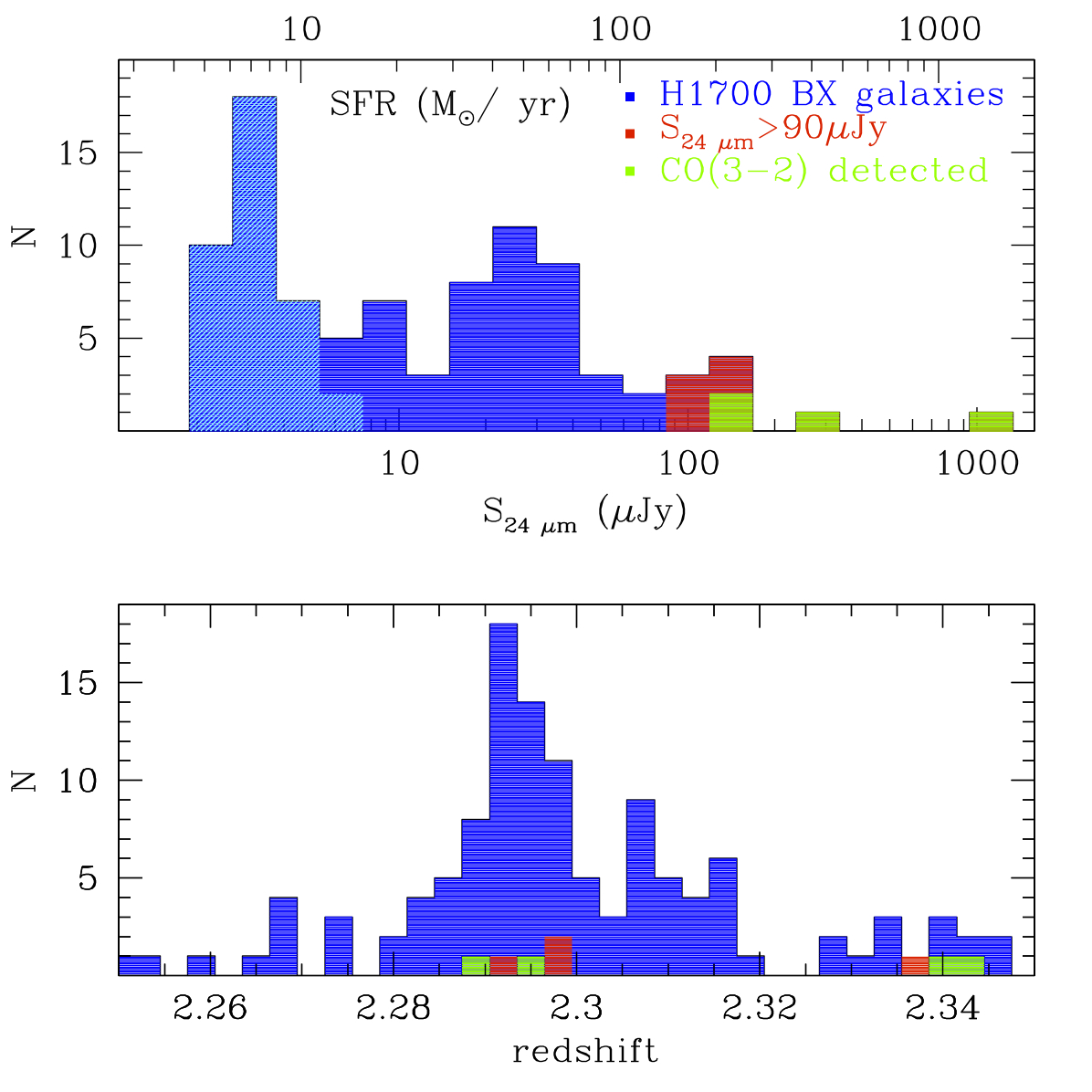}
\caption{{\bf Left  Top:}  Multi-wavelength SED of HS1700.DRG55, compared to the ALMA ALESS SMG composite (Swinbank et al.\ 2014) scaled to a 2mJy 850$\mu$m source at $z=2.3$. DRG55 is remarkably similar to the source, but optically fainter shortward of $1\mu$m.
{\bf Left Bottom:} Comparison of the optical/IR colours of DRG55 (red symbol)   with  spectroscopically confirmed UV-selected galaxies in the same field from Shapley et al.\ (2005),  emphasizing the redness of the source relative to typical UV galaxies. The Tacconi et al.\ (2013) CO-detected BX galaxies are also highlighted with larger circles.
{\bf Middle:}   SFR versus L$'$CO   for various literature galaxies compared with DRG55 (blue circles - Tacconi et al.\ 2013; red squares - Bothwell et al.\ 2013). The solid magenta pentagons show L$'$(CO) for the 24$\mu$m-brightest galaxies in  our survey, with limits at 3$\sigma$.  DRG55 is nominally over-luminous in $^{12}$CO compared to other ultra-luminous galaxies, but similar to  other  Tacconi et al.\ (2013) LBGs lying in the proto-cluster (larger blue circles). 
{\bf Right Top:}  The 24$\mu$m fluxes of all proto-cluster galaxies with spectroscopic redshifts, with inferred SFR labelled on top axis. The brightest 24$\mu$m sub-sample, and those that are detected in CO(3--2) are highlighted. For S$_{\rm 24\mu m} < 10\mu Jy$ (2$\sigma$ limit of the {\it Spitzer}-MIPS map), we have estimated the 24$\mu$m fluxes from the UV-derived SFR (cyan shading). 
{\bf Right Bottom:} Histogram of the proto-cluster galaxies, with the same sub-populations highlighted. The 24$\mu$m-bright, and CO-detected sub-samples appear to span a larger range in redshifts than the the overall cluster population.
}
\label{spectra} 
\end{figure*}

\section{Observations}

We have obtained an observation centered at 100.4 GHz in the 3-mm window performed at PdBI at a 
pointing center: RA=17 01 17.62,     Dec=+64 14 38.0  (J2000.0), 
designed to perform a blind census of $^{12}$CO(3--2) near the $z=2.300$ proto-cluster core, and specifically targeted a bright Sub-Millimeter Galaxy (SMG) thought to potentially lie at $z$=2.31  given its location 2$''$ from an extended complex of UV-luminous star forming galaxies 
(the SMG turned out to have $z=2.82$ -- S.\ Chapman et al.\ in prep).
The galaxy DRG55 described in this Letter lies 9.1$''$ from the pointing centre.
Observations were obtained on April 13 and 14, 2014, over two tracks (project S04CK) in D configuration, using the  5-antenna sub-array.
We used the PdBI WideX correlator covering frequency range of 3.6GHz (102.1GHz -- 105.7GHz), corresponding to redshifts $2.271<z<2.387$ in the $^{12}$CO(3--2) 
line. 
 Flux calibration 
 used 3C273 and 3C345, and the quasar B1749+096 was used as a phase and amplitude calibrator.  
We resampled the datacubes in 60 km s$^{-1}$ channels, and imaged them using the {\sc gildas} suite {\sc mapping} using natural weighting. 
The  beam size is 6.6$''$$\times$4.1$''$, PA=61.7 deg ($\sim$30$\times$45 kpc at $z$=2.3). 
The final cube 
corresponds to 3.34 hr on source. 
To obtain flux measurements, we deconvolved the visibilities using the {\sc clean} task with natural weighting, and applied the corresponding primary beam correction. 

%

The  data has an RMS of 0.6 mJy beam$^{-1}$  in  50 MHz 
channels., corresponding to a 3.5$\sigma$  limit in 
 $^{12}$CO  of 0.4~Jy km s$^{-1}$  or L$_{\rm lim}$ = 1.8$\times10^{10}$ K km s$^{-1}$ pc$^2$ at $z\sim2.30$ (assuming a typical line width of 300 km s$^{-1}$, 0.38 mJy RMS,  and an average primary beam correction of 0.72).
This is comparable	with   faint $^{12}$CO detections reported  from 3mm observations at these redshifts (e.g.\ Bothwell et al.\ 2013; Tacconi et al.\ 2013).

Submillimeter Array (SMA) interferometric imaging observations at 870$\mu$m were also taken in this field, using both compact and extended configurations.  The SMA $\sim$35$''$ primary beam at this wavelength encompasses both the original targeted SMG as well as the 9$''$ offset position of DRG55, where the primary beam correction is 15\%. The combined map reaches an RMS of 1.0mJy at phase center (Chapman et al. in prep) with a resolution of  $\sim$1$''$.
The SCUBA-2 map of this field, reaching the confusion limit with an RMS at 850$\mu$m of 0.5\,mJy is also described elsewhere (Chapman et al.\ in prep.).

\subsection{The four archival proto-cluster PdBI pointings} 

Archival observations of $^{12}$CO emitters by  Tacconi et al.\ (2013) observed four Lyman-Break Galaxies (LBGs)  lying within the proto-cluster at $z=2.3$ using the old generation receivers (giving velocity coverage of $\sim$2000\ km s$^{-1}$, detecting three of them robustly in $^{12}$CO(3--2). Despite the arguments in Tacconi et al.\ (2013) for these being {\it normal} star forming galaxies, all four actually have AGN-like near-IR spectra (from [N{\sc II}]/H$\alpha$-- Erb et al.\ 2011), while MD174 is clearly an AGN even in its UV spectrum. A SCUBA-2 850$\mu$m survey of these CO-observed galaxies   (Beanlands et al.\ in prep) shows that on average these sources have S$_{\rm 850 \mu m}$=1.4\,mJy, although a few (in fields other than HS1700+64) have substantially higher S$_{\rm 850 \mu m}\sim4$\,mJy.
 
As we detect DRG55 in $^{12}$CO within 1000 km s$^{-1}$ of our targeted z=2.305, there is some hope that the Tacconi PdBI observations might uncover additional proto-cluster sources offset from the targeted LBG. 
We retrieved the Tacconi et al.\  datacubes from the archive and searched  for blind $^{12}$CO detections offset from the targeted  galaxies.
We confirm their three of four $^{12}$CO detections of the BX galaxies, but do not find any additional $^{12}$CO sources to depths similar to our present study. 
In addition there are no known optical galaxies with redshifts in the $^{12}$CO range in any of these four pointings (summary in Table~1), although two lie very near the frequency band edge in the MD69 and MD174 pointings.

\section{Results} \label{S:results}

Our PdBI observations 
cover the $^{12}$CO(3-2) line emission from nine galaxies (listed in Table~1) near the proto-cluster core that either have an  optical spectroscopic redshift in the range $2.28<z<2.38$ (five galaxies) or appear luminous at 24$\mu$m and have colors consistent with $z\sim2.3$ (three DRGs and an SMG).  They allow for initial  
exploration of the diversity of gas-rich galaxies in the proto-cluster, and to quantify the potential missing sources.
 %
 This field represents a particularly overdense sub-region of the proto-cluster, triple the average overdensity over 20 arcmin$^2$, and
 lying $\sim1.5'$ from the optically defined proto-cluster core. 

Since the frequency of the $^{12}$CO(3-2) line emission could be slightly offset from the frequency implied by the optical redshifts, we 
searched for a $^{12}$CO emission peak in the spectra along the line of sight. 
Only DRG55 shows a clear line detection (although the SMG at pointing centre is  detected $>5\sigma$ in 3-mm continuum at 0.380$\pm$0.075\,mJy).
When we stack the seven non-detections, we do not find any significant line emission, but set an improved limit of 
L$_{\rm lim, 3\sigma}$ = 0.8$\times10^{10}$ K km s$^{-1}$ pc$^2$ on their average $^{12}$CO(3--2), the average expected emission from a 100 M$_\odot$\,yr$^{-1}$ starburst (Fig.~2, and Tacconi et al.\ 2013)

An optimal extraction of DRG55 
was found by summing channels from 104.76 to 105.15 GHz 
($\sim$1100 km s$^{-1}$), showing a 4.8$\sigma$ detection of 1.1$\pm$0.2\,mJy at the position of DRG55 (Fig.~1), coincident (to 0.2$''$) with a 
160$\mu$Jy 24$\mu$m source.  There are no other peaks in the channel map this strong: the next two are $3.5\sigma$ and $2.8\sigma$.
The extracted 1D spectrum shows a well detected $^{12}$CO(3--2) line peaking at $\sim$2\,mJy  (Fig.~1) with a continuum RMS=0.23\,mJy, and a line flux of 1.14$\pm$0.31\,Jy\,km\,s$^{-1}$, and an implied line luminosity of $3.6\times10^{10}$ K\,km$^{-1}$\,pc$^2$.
The total molecular gas mass of $3.2\times10^{11}$ M$_\odot$ is derived, corrected for Helium (1.36), a ``Galactic'' CO--H$_2$ conversion factor  $\alpha$ = 4.36, and a CO(1--0)/CO(3--2) ratio of two, systematic uncertainty 50\%.  
 A Gaussian fit to the line suggests a peak frequency of 104.9\,GHz, and $z_{\rm CO}=2.296$, near the central proto-cluster redshift of $z=2.300$ (Steidel et al.\ 2005).

While the SCUBA-2 map cannot easily place a limit on the 850$\mu$m emission due to the bright SMG $10''$ away, it shows no detection at 450$\mu$m (7$\pm$6\,mJy) where the beam is 7$''$. The SMA map 
shows an 850$\mu$m emission at the position of DRG55 of 2.5$\pm$1.1\,mJy or  $<3.3$mJy ($3\sigma$).
DRG55 is essentially undetected in optical ground-based imagery ($R\sim27$, $g>27.3$, $U>27.2$, $3\sigma$). 
 In an {\it HST} ACS F814W image (details in Peter et al.\ 2007), DRG55 is identified with a faint $I_{\rm AB}=26.0$ galaxy with {\it tadpole} morphology (Fig.~1), and is well detected in near/mid-IR ($J_{\rm AB}=23.92,  K_{\rm AB}=22.30,$ {\rm IRAC}-$ch2_{\rm AB}=21.9$).
 
Our constraint on the FIR  emission is really grounded by the S(24$\mu$m)=0.16$\pm$0.01\,mJy point, and the S$_{\rm 850 \mu m}$$<$3mJy limit. However the 24$\mu$m point sits at the peak of the 7.7$\mu$m PAH line, and as such is also subject to systematic uncertainty.
Nonetheless, from IRAC through submm wavelengths, the SED is remarkably similar to the Swinbank et al.\ (2014) composite SMG SED from ALESS (Fig.\ 2), from which we derive an L$_{\rm FIR}=2.1\times10^{12}$ L$_\odot$ and a SFR$\sim$210 M$_\odot$ yr$^{-1}$ (Kennicutt 1998, Chabrier IMF). 
The peak in the SED at IRAC 4.5$\mu$m  suggests there is no obvious AGN, and the 24$\mu$m emission may therefore be SF powered, contrasted with the  AGN-like galaxies hosting the other three known $^{12}$CO sources in the proto-cluster (as described in \S~2.1).

A Gaussian fit to the $^{12}$CO line of  FWHM 680$\pm$141 km\,s$^{-1}$, together with a size of 0.35$''$ (half-light radius) in the HST F814W image, implies a dynamical mass of 3.7$\times10^{11}$ M$_\odot$, assuming a virial mass indicator. The $J,K$, and IRAC 4.5$\mu$m luminosities suggest a stellar mass of  $\sim$1$\times10^{11}$ M$_\odot$.   The dynamical constraint thus sets a limit to the gas mass of $<$2.7$\times10^{11}$ M$_\odot$. 
Another approach is to estimate  the gas fraction  as m$_{\rm gas}$/(m$_{\rm stars}$ + m$_{\rm gas}$), suggesting a very gas rich system of  $f_{\rm gas}\sim$80\% with $\alpha_{\rm CO}$=4.36, although the stellar mass is uncertain with systematic error of a factor two (e.g., Hainline et al.\ 2011). For high$-z$ SMGs, a $\sim$50\% gas fraction was estimated  (Bothwell et al.\ 2013), but using a lower $\alpha$=1-2. 
If we adopt similarly low $\alpha_{\rm CO}$, $f_{\rm gas}$ is $\sim$50\% for DRG55, in line with the ULIRGs from Bothwell et al.\ (2013).

To search for any additional proto-cluster members which may not be associated to known optical galaxies, 
we also performed a blind search for significant emission line peaks in the PdBI data cube, lying within the primary beam.
We performed the search making use of the {\sc AIPS} task {\sc SERCH}, which uses a Gaussian kernel to convolve the data cube along the frequency axis with an expected input line width, and reports all channels and pixels having a signal to noise ratio over the specified limit. We experimented with  Gaussian kernel line widths, from $\sim260$ to 520 km s$^{-1}$. We found no peaks at $>5\sigma$ ranging over 10,000 km s$^{-1}$ in velocity space. At $>4\sigma$, 
we find an additional three candidate lines ranging from 4.4$\sigma$ to 4.7$\sigma$, but all with velocity offsets $>2000$km s$^{-1}$ from $z=2.300$,  and with distances from the pointing center of 10.2$''$ to 24.3$''$.  None of these lie within $<2''$ of any IRAC 4.5$\mu$m sources, and may be spurious lines, but in any case are distant enough in velocity to be outliers from the proto-cluster.








\begin{table*}
\label{tab:par}
 \flushleft{
  \caption{CO-targeted galaxies within the Primary Beam of our survey (upper entries), and summary of Tacconi et al.\ (2013) $^{12}$CO-observed BX galaxies in the same proto-cluster (lower entries).
  Submm fluxes are either 850$\mu$m from SCUBA-2, with the $\sim$0.5mJy RMS listed, or 870$\mu$m from SMA followup (where the SCUBA-2 map cannot be used to set useful limits due to the bright SMG).
  Molecular gas mass, corrected for Helium, a ``Galactic'' CO--H$_2$ conversion factor, $\alpha$ = 4.36, and a CO(1--0)/CO(3--2) ratio of 2.
  The last four galaxies were selected by the Lyman-Break Galaxy colour criteria (although all four actually have AGN-like near-IR spectra), with 850$\mu$m fluxes 
  from J.\ Beanlands et al.\ (in prep). The DRG55 redshift is obtained solely from $^{12}$CO(3-2).}
\begin{tabular}{llccccccc}
\hline
ID  &   RA / Dec (J2000)& Rad.\ &  $z$ & S$_{\rm 850 \mu m}$  & S$_{\rm 24 \mu m}$ & SFR & M* & M$_{\rm g}$ \\ 
{}  &   {}  & ({$''$}) & {} & (mJy) & ($\mu$Jy) & {M$_\odot$ yr$^{-1}$}& {$\times$10$^{10}$ M$_\odot$}  & {$\times$10$^{11}$ M$_\odot$}\\ 
\hline
BX951 	   & 17\,01\,17.90     +64\,14\,40.5 & 2.9 & 2.308 & $<3$ & $<$10 & $<$13 & 3.8 & $<1$ \\  
DRG55    & 17\,01\,19.37  +64\,14\,37.7   &9.1  & 2.296 & $<3$ & 160.1$\pm$5.0  & 210 & 9.8 & 3.2 \\ 
BX917	& 17\,01\,16.15     +64\,14\,19.6 & 20.1 &  2.304 & $<3$ & 110.9$\pm$5.0  & 146 & 4.5 &$<2$\\ 
DRG71    &   17\,01\,16.49    +64\,14\,15.4 & 22.3 & $\sim$2.3$\pm$0.2 & $<3$ & $<$10 & $<$13 & 2.0 &$<2$\\ 
DRG53 & 17\,01\,21.06     +64\,14\,22.3 & 23.3 & $\sim$2.3$\pm$0.2 & 1.7$\pm$0.5 & 604$\pm$5.0  & 793 & 75.7 &$<2$\\  
BX950   & 17\,01\,14.19     +64\,14\,40.4 & 25.2 & 2.311& $<3$ &  $<$10 & $<$13 & 1.3 &$<2$\\ 
DRG73& 17\,01\,21.75     +64\,14\,36.0 & 25.4& $\sim$2.3$\pm$0.2  & $<3$ & $<$10 & $<$13 & $<0.9$ &$<2$\\ 
BX913      	& 17\,01\,21.61     +64\,14\,18.2  &  29.2 & 2.291 & 2.1$\pm$0.5 & 30.2$\pm$5.0  & 40 & 4.0 &$<2$\\ 
\hline
MD103$^1$  & 17\,01\,00.25     +64\,11\,55.3 & {} &  2.315 & $<1.5$ & 164.2$\pm$5.0 & 216 & 6.6 &$<$1.1 \\ 
MD174$^2$  & 17\,00\,54.58     +64\,16\,24.5 & {} &2.344 & 1.7$\pm$0.5 & 1116.5$\pm$5.0 & 1466 & 24.0 &1.7 \\ 
MD69$^3$  & 17\,00\,47.65     +64\,09\,44.5 &{} &2.288 & 1.8$\pm$0.5 & 256.8$\pm$5.0 &  337 & 19.0 &0.96 \\
MD94$^4$  & 17\,00\,42.06     +64\,11\,24.0 &{} &2.340 & 1.3$\pm$0.5 & 57.3$\pm$5.0 & 75 & 15.0 &3.3 \\ 
\hline
\end{tabular}\\
\footnotemark{MD103 is listed as undetected  in $^{12}$CO(3--2) in Tacconi et al.\ (2013), consistent with our analysis. All other known LBGs in the PB are at $z\sim2.75$. }
\footnotemark{MD174 is  a strong AGN  from UV/near-IR spectra. 
Another BX1156 $z=2.285$, lies just  beyond $z=2.327-2.357$ covered by PdBI. A  bright S$_{\rm 850 \mu m}$=7\,mJy SMG  lies  within the PB, but no lines are detected: we conclude it is not in the covered redshift range, but could still be in proto-cluster.}\\
\footnotemark{Another galaxy, BX505 z=2.309, is just beyond (545 km s$^{-1}$) the  $z=2.273-2.303$ probed by  PdBI. The galaxy is undetected in $^{12}$CO at band edge.}\\
\footnotemark{All other known galaxies with spectroscopic redshifts falling in the PB lie outside the redshift range $z=2.325-2.355$.}
}
\end{table*}

%




\section{Discussion and Conclusions} \label{S:discussion}

The detection of DRG55 in $^{12}$CO complements three other $^{12}$CO-detected UV-bright galaxies in this proto-cluster from Tacconi et al.\ (2013),  suggesting that many optically faint galaxies in the proto-cluster may host substantial molecular gas reservoirs, and a full blind census of $^{12}$CO in this overdense environment is warranted.
To assess how many additional $^{12}$CO-bright sources might be found  in this region,
we look first at  the likely size of such a proto-cluster
%
(the pieces which end up within the central 1~Mpc at $z\sim0$), 
for example Chiang et al.\ (2013) 
predict 5-10 comoving Mpc.  While we have targeted an overdense sub-region within the proto-cluster, our PdBI survey size of only 0.5~Mpc 
 in diameter. While the full proto-cluster may well extend over 100-200 times the area, less than $\sim$one tenth will be as overdense as this DRG55 sub-region (already three times the average core density). 
We thus might expect that in a blind survey in this proto-cluster, we would uncover 10-20 such $^{12}$CO(3--2) emitters.
Under the assumption that Tacconi et al.\ (2013) fairly sampled the top end of the UV-bright luminosity function in the proto-cluster (Fig.~2), a similar number ($\sim20$) of UV-bright galaxies with L$'$(CO)$>1\times10^{10}$~K\,km\,s$^{-1}$\,pc$^2$ should reside in the full extent of the proto-cluster. 
 
A color-luminosity plot (Fig.~2) for the proto-cluster population is used to compare DRG55 with the Tacconi et al.\ (2013) galaxies and the general $z\sim2.3$ UV-population. DRG55 is certainly significantly fainter and redder, and suggests that there may well be other gas rich members of the proto-cluster that are not accounted for by the UV-bright census.
We learn that the Tacconi et al.\ galaxies are amongst the reddest in the population, but that DRG55 still separates significantly in optical properties despite having  a similar $^{12}$CO mass. 
DRG55 is one of 72 near-IR selected objects in the surrounding  9$'$ $\times$ 9$'$ field, satisfying the  ($J$-$K_s$)$_{\rm vega}$$>$2.3 
    criteria advocated by Franx et al.\ (2003) for selecting $z$$>$2 red galaxies.  
    DRG55 is significantly fainter in the rest-UV than typical 
    LBG-selected objects, but its near-IR properties are not particularly unusual. 
 It is one of only two DRGs in the 9$'$ field that is also flagged as an H$\alpha$ narrow-band candidate (which was designed to probe
      the proto-cluster redshift -- Erb et al.\ 2014), setting some  limit on the number of additional objects with similar near-IR-bright properties that could be found within the $z\sim2.3$ structure.   
  Its  H$\alpha$+[NII] flux is 8$\times10^{-17}$ ergs\,s$^{-1}$\,cm$^{-2}$\,Hz$^{-1}$ (relatively easy to reach 
  with near-IR spectroscopy), corresponding to a SFR of $\sim$15 M$_\odot$ yr$^{-1}$ before
     applying an extinction correction (a factor 14$\times$ to match the SFR$_{\rm 24\mu m}$).  
     The H$\alpha$ line excess 
also means that the $^{12}$CO  redshift is almost certainly $z$=2.296. 
 DRG55 is not detected in deep Lyman-$\alpha$ images of the field reaching  
 $<1\times10^{-17}$ ergs\,s$^{-1}$\,cm$^{-2}$\,Hz$^{-1}$ 
(Erb et al.\ 2011; 
Steidel et al 2011). 
We also put DRG55 in the context of other photometrically studied populations,
specifically the `dust-obscured' galaxies with high f$_{24\mu m}$/f$_{R}$$>$1000 (e.g.\ Dey et al. 2008, Penner et al. 2012),  for which DRG55 qualifies, 
although galaxies selected in this way are a mixture of dusty star formers and  
obscured AGN. $^{12}$CO observations exist only for few rather luminous similar sources (Yan et al.\ 2010). Interesting $^{12}$CO-detected (though more luminous) comparison objects may also be obscured SMGs such as SMM14009 (Weiss et al.\ 2009) or SMMJ00266+1708 (Sharon et al.\ 2015), and several from Chapman et al.\ (2005).
          A targeted search for $^{12}$CO in the proto-cluster around faint DRGs identified to 24$\mu$m or 850$\mu$m sources, with phot-$z$'s consistent with $z\sim2.30$ (but too faint for optical spectroscopy) might quickly increase the total number of known ultra-luminous cluster members. 

We next turn to a discussion of DRG55 versus other 
$^{12}$CO-detected sources, 
plotting L$'$(CO)-SFR for the high-$z$ population (Fig~2). 
DRG55 is somewhat over-luminous in $^{12}$CO for its SFR compared to SMGs from Bothwell et al.\ (2013), as are two of the Tacconi  et al.\ (2013) proto-cluster $^{12}$CO sources, suggesting a low-efficiency of star formation.  
Our survey limit is  sufficient to detect the average relation for a SFR$\sim$200 M$_\odot$\,yr$^{-1}$, but can only detect over-luminous $^{12}$CO emitters at  lower SFRs. 
%
Fig.~2 also highlights the 24$\mu$m properties of all known proto-cluster galaxies, revealing that DRG55 has  a similarly high IR luminosity to the Tacconi et al.\ (2013) galaxies (although one of their sources is a clear outlier with S$_{\rm 24\mu m}$$\sim$1\,mJy). 
%
To date, the only truly blank field, blind high-$J$ CO survey conducted is that of Decarli et al.\ (2014) 
who found two secure CO detections in a deep PdBI scan of the HDF-N (their ID.03 \& ID.19), which they identified with star-forming galaxies at $z$=1.784 and at $z$=2.047. 
They found these galaxies to have colors consistent with the BzK galaxies (Daddi et al.\ 2008),  although with specific SFRs below the locus of $z\sim2$ main-sequence galaxies. They show relatively bright $^{12}$CO emission compared with galaxies of similar dust continuum luminosity (but still within the observed scatter of the L$_{IR}$--L$'$(CO) relation).



Finally we assess the redshift and spatial distribution of DRG55 relative to the other known $^{12}$CO emitters in the proto-cluster.
Of note is that the $^{12}$CO-detected galaxies  comprise the four brightest 24$\mu$m sources known to be in the proto-cluster, with estimated SFRs(24$\mu$m) ranging from 200-1100 M$_\odot$\, yr$^{-1}$ (Fig.~2).
There are of course many potential dust-obscured SMGs lying in the proto-cluster, yet to be identified. None of the SCUBA-2 sources in this field have spectroscopic redshifts, but many have robust DRG-counterparts which have colors consistent with lying in the proto-cluster.
The redshifts of these four $^{12}$CO emitters divide into two redshift groups, also mirrored by the eight brightest 24$\mu$m emitters known in the proto-cluster (Fig.~3). One  group is in the central proto-cluster `spike', with the second group in a knot of 
more receding objects, which might represent  a gas-rich group falling in from in front and being lit up. However, this background `group' does not cluster spatially within the $9'$ field, and may be more {\it sheet}-like.  These eight galaxies have a much broader dispersion than the proto-cluster redshift distribution, and may indicate environmental triggering of the most luminous population.
\vskip-2cm
\acknowledgments
This work is based on observations carried out with the IRAM Plateau de Bure Interferometer. IRAM is supported by INSU/CNRS (France), MPG (Germany) and IGN (Spain). 
The Submillimeter Array is a joint project between the  Smithsonian Astrophysical Observatory and the Academia Sinica Institute of Astronomy and Astrophysics and is funded by the Smithsonian Institution and the Academia Sinica.
SC and IS acknowledge the Aspen Center for Physics where parts of this manuscript were written.
SC acknowledges support from NSERC and CFI.

\end{document}